\documentclass[aps,manuscript,reprint,noshowpacs,noshowkeys,prd,balancelastpage,nofootinbib]{revtex4}
\usepackage[utf8]{inputenc}
\usepackage{color}
\usepackage{amsmath}
\usepackage{amssymb}
\usepackage[unicode=true,
 bookmarks=false,
 breaklinks=false,pdfborder={0 0 1},backref=section,colorlinks=true]
 {hyperref}
\hypersetup{
 linkcolor=blue,urlcolor=blue,citecolor=red}

\makeatletter
\@ifundefined{textcolor}{}
{%
 \definecolor{BLACK}{gray}{0}
 \definecolor{WHITE}{gray}{1}
 \definecolor{RED}{rgb}{1,0,0}
 \definecolor{GREEN}{rgb}{0,1,0}
 \definecolor{BLUE}{rgb}{0,0,1}
 \definecolor{CYAN}{cmyk}{1,0,0,0}
 \definecolor{MAGENTA}{cmyk}{0,1,0,0}
 \definecolor{YELLOW}{cmyk}{0,0,1,0}
}

\usepackage{amsfonts}
\usepackage{mdframed}
\usepackage{footnote}
\usepackage{float}
\usepackage[font=footnotesize,it]{caption}
\usepackage{tikz}
\usepackage{tikz-3dplot}

\makeatother

\begin{document}
\title{Generalized probability and current densities: A field theory approach}
\author{M. Izadparast }
\email{masoumeh.izadparast@emu.edu.tr}

\author{S. Habib Mazharimousavi}
\email{habib.mazhari@emu.edu.tr}

\affiliation{Department of Physics, Faculty of Arts and Sciences, Eastern Mediterranean
University, Famagusta, North Cyprus via Mersin 10, Turkey}
\date{\today }
\begin{abstract}
We introduce a generalized Lagrangian density - involving a non-Hermitian
kinetic term - for a quantum particle with the generalized momentum
operator. Upon variation of the Lagrangian, we obtain the corresponding
Schrödinger equation. The extended probability and particle current
densities are found which satisfy the continuity equation.
\end{abstract}
\maketitle

\section{Introduction}

The idea of the generalized momentum operator has been extended in
our earlier proposal \cite{M.H1,M.H2}, which is subjected to the
non-relativistic and non-Hermitian quantum mechanics. The thought
of the non-Hermitian Hamiltonian was discussed initially in Ref. \cite{Bessis,Caliceti}.
Yet, Bender and Boettcher have initiated $\mathcal{PT}$-symmetric
quantum physics by employing a harmonic-oscillator-like $\mathcal{PT}$-symmetric
potential as a toy model \cite{Bender1}. They have shown that even
if the Hamiltonian is non-Hermitian the eigenvalues can be real, upon
which, a $\mathcal{PT}$-symmetric operator is defined. The $\mathcal{PT}$
operator is composed of the parity and the time reversal operators,
namely i. e., $\mathcal{P}x\mathcal{P}=-x$ and $\mathcal{T}i\mathcal{T}=-i$
\cite{Bender2}. For a $\mathcal{PT}$-symmetric Hamiltonian - a less
general non-Hermitian Hamiltonian - the interaction potential is considered
to be complex. Hence, the unitarity for the unbroken $\mathcal{PT}$-symmetry
(under which the energy spectrum of the system is real) in a quantum
mechanical system has to be conserved, which is one of the prominent
principles of quantum mechanics. Accordingly, in Ref. \cite{Bender3}
the infinitesimal probability density has been considered and a method
introduced to find the path integral in the complex plane $\mathbb{C}$.
For further investigations on the presented outcomes in Ref. \cite{Bender1},
the approach of field theory, also, is presented in the context of
$\mathcal{PT}$-symmetric quantum physics. Bender et. al. in the field
approach have discussed the idea of broken symmetry of a separated
parity and time reversal operators, although the field is $\mathcal{PT}$-symmetric,
in Ref. \cite{Bender8}. Over the latter study, the idea has been
expanded in a two-dimensional supersymmetric quantum field theory
according to a defined potential $-ig\left(i\phi\right)^{1+\delta}$
for $\delta>0$. Besides, in Ref. \cite{Bender7}, the authors have
applied the technique of truncating the Schwinger-Dyson equations
in a set of fields with the non-Hermitian Hamiltonian families, $\frac{g}{N}\left(i\phi\right)^{N}$,
provided that $N\geq2$. In the latter paper, the corresponding solution
and renormalization have been obtained and the properties of the scalar
quantum field $-g\phi^{4}$ in four-dimensional space-time discussed.
Later on, in Ref. \cite{Bagchi}, Bagchi et. al. have argued meticulously
the $\mathcal{PT}$-symmetric field theory upon the variation of Lagrangian
to find the generalized continuity equation, using the Schrödinger
equation and its $\mathcal{PT}$-symmetric conjugate. Furthermore,
the modified normalization constant has been obtained on the real
$x$-axis. A significant study has been carried out in Ref. \cite{Bender4}
in which presents a perturbative method to find the $\mathcal{C}$
operator in quantum mechanics, included systems with higher degrees
of freedom, and particularly in quantum field theory. The same authors
in a short letter in Ref. \cite{Bender5} have presented the successful
physical confirmation of $\mathcal{PT}$-symmetric quantum field theory
corresponding to the field $i\phi^{3}$, and redefined the inner product
in Hilbert space using the perturbation method to build $\mathcal{C}$
operator. The Hamiltonian of a free fermionic field theory in Ref.
\cite{Bender6} has been investigated and shown that its $\mathcal{PT}$
invariance depends on the corresponding mass term. According to a
study in Ref. \cite{Bender6}, it has been confirmed that the $\mathcal{PT}$-symmetric
massive Thirring and the scalar sine-Gordon models are dual to each
other and equivalent to their Hermitian version. Regarding the importance
of field $\phi^{3}$ in the context of $\mathcal{PT}$-symmetric quantum
field theory, in Ref. \cite{Bender11} the authors have compared the
renormalization-group properties of Hermitian field $g\phi^{3}$ and
$\mathcal{PT}$-symmetric $ig\phi^{3}$ field theories. However, in
Ref. \cite{Bender12}, the critical behavior of $\mathcal{PT}$-symmetric
$i\phi^{3}$ quantum field theory has been studied in $6-\epsilon$
dimensions around the exceptional points. Furthermore, the calculation
on the the critical exponent has been carried out employing the mean-field
approximation and the renormalization-group technique. In Ref. \cite{Bender10},
having considered $\mathcal{PT}$-symmetric quantum theory, the logarithmic
time-like Liouville Lagrangian has been discussed. Accordingly, the
authors have found the energy of a quantum mechanical system assuming
the semiclassical limit. Recently, Alexandre et. al. in a remarkable
study in Ref. \cite{Alexander} have argued Noether theorem considering
complex scalar and fermionic $\mathcal{PT}$-symmetric field theories
for Hermitian and Anti-Hermitian mass using the variation of Lagrangian.
Very recently, Mazharimousavi rigorously, in Ref. \cite{Mazhari}
through the approach of classical field theory, has declared that
one field is sufficient to find a nonlinear or generalized Schrödinger
equation in the context of the position dependent mass (PDM) or deformed
momentum operator. Besides, in Ref. \cite{Mazhari}, it has been indicated
that the generated field equations lead to the standard Schrödinger
equation. Whereas, the corresponding probability density remains real
and positive in both Hermitian and non-Hermitian approaches

Here in this paper, we introduce a generalized Lagrangian density
extending the latter approach, in Ref. \cite{Mazhari} through the
conventional Hermitian or $\mathcal{PT}$-symmetric field theory using
the generalized momentum operator \cite{M.H1,M.H2}. We study the
physical aspects of the Lagrangian density including the corresponding
Schrödinger equations and the continuity equation.

The present paper is organized as follows. In Sec. II we introduce
the generalized Lagrangian density which corresponds to the extended
definition of the generalized momentum operator in the real domain.
Besides that, by applying the principle of stationary action through
the Euler-Lagrange equation, we obtain the generalized Hamiltonian
and the corresponding Schrödinger equation. Then, the continuity equation
is found under the concept of the generalized probability and current
densities. In Sec. III and Sec. IV we discuss the Lagrangian densities
in the complex domain once for the conventional momentum operator
and, again, for the generalized momentum operator, respectively. Finally,
we summarize our paper in the Conclusion.

\section{Lagrangian Density in Real Domaine: Hermitian Schrödinger Equation}

We recall the generalized momentum operator presented in Ref. \cite{M.H1,M.H2}
or in the general form in Eq. (10) in Ref. \cite{Mazhari}, i.e 
\begin{equation}
\hat{p}=-i\hbar\left(A\partial_{x}+\frac{A^{\prime}}{2}\right),\label{generalized-momentum}
\end{equation}
 where the auxiliary function $A\left(x\right)$ is a real function
of $x$. Then, we introduce the generalized Lagrangian density 
\begin{equation}
\mathcal{\mathcal{L}=}i\hbar\dot{\Psi}\Psi^{*}-\frac{\hbar^{2}}{2m}A^{2}\Psi^{*\prime}\Psi^{\prime}-\left[-\frac{\hbar^{2}}{4m}AA^{\prime\prime}-\frac{\hbar^{2}}{8m}A^{\prime2}+V\left(x\right)\right]\Psi\Psi^{*},\label{Lagrangian-generalized}
\end{equation}
which a dot and prime stand for the derivative with respect to $t$
and $x$, respectively, and {*} implies the complex conjugate. By
Applying the variation of the action with respect to $\Psi^{*}\left(x,t\right)$
and $\Psi\left(x,t\right)$ and choosing the interaction potential
$V\left(x\right)$ to be real the field equations are expressed as

\begin{equation}
i\hbar\dot{\Psi}=-\frac{\hbar^{2}}{2m}\left(\left(A^{2}\Psi^{\prime}\right)^{\prime}+\frac{A^{\prime\prime}A}{2}\Psi+\frac{A^{\prime2}}{4}\Psi\right)+V\left(x\right)\Psi\label{Schrodinger-psi}
\end{equation}
and
\begin{equation}
-i\hbar\dot{\Psi}^{*}=-\frac{\hbar^{2}}{2m}\left(\left(A^{2}\Psi^{*\prime}\right)^{\prime}+\frac{A^{\prime\prime}A}{2}\Psi^{*}+\frac{A^{\prime2}}{4}\Psi^{*}\right)+V\left(x\right)\Psi^{*},\label{Schrodinger-phi}
\end{equation}
respectively. Eqs. \eqref{Schrodinger-psi} and \eqref{Schrodinger-phi}
give the generalized Schrödinger equation and its complex conjugate.
To find the stationary form of latter equations, we use the field
in terms of separable time and spatial functions given by 
\begin{equation}
\Psi\left(x,t\right)=\psi\left(x\right)\exp\left(-\frac{iE}{\hbar}t\right),\label{field}
\end{equation}
in which plugging into the field equation \eqref{Schrodinger-psi}
leads to 
\begin{equation}
E\psi=-\frac{\hbar^{2}}{2m}\left(\left(A^{2}\psi^{\prime}\right)^{\prime}+\frac{A^{\prime\prime}A}{2}\psi+\frac{A^{\prime2}}{4}\psi\right)+V\left(x\right)\psi.\label{motion}
\end{equation}
Whilst, the Hamiltonian operator $\hat{H}$ in Ref. \cite{M.H1,M.H2}
is defined to be 
\begin{equation}
\hat{H}=\hat{H}^{\dagger}=-\frac{\hbar^{2}}{2m}\left(A^{2}\partial_{x}^{2}+2AA^{\prime}\partial_{x}+\frac{A^{\prime\prime}A}{2}+\frac{A^{\prime2}}{4}\right)+V\left(x\right),\label{Original-Ham}
\end{equation}
which is accordingly admits Eq. (10) in Ref. \cite{Mazhari}. Next,
we use the Lagrangian density \eqref{Lagrangian-generalized} and
the definition of Hamiltonian density 
\begin{equation}
\mathcal{H}=\Sigma_{\sigma}\pi_{\sigma}\dot{f_{\sigma}}-\mathcal{L},\label{Hamiltonian-den}
\end{equation}
 where $f_{\sigma}\in\left\{ \Psi,\Psi^{*}\right\} $ and $\pi_{\sigma}$
is the momentum-density conjugate to $f_{\sigma}$, and calculate
the explicit form of $\mathcal{H}$. To do so, let's calculate $\pi_{\sigma}$
as 
\begin{equation}
\pi_{\Psi}=\frac{\partial\mathcal{L}}{\partial\dot{\Psi}}=i\hbar\Psi^{*},\,\pi_{\Psi^{*}}=\frac{\partial\mathcal{L}}{\partial\dot{\Psi}^{*}}=0.\label{canonical momenta}
\end{equation}
After the substitution into Eq. \eqref{Hamiltonian-den}, one-dimensional
Hamiltonian density is obtained to be 
\begin{equation}
\mathcal{H}=\frac{\hbar^{2}}{2m}A^{2}\Psi^{*\prime}\Psi^{\prime}-\frac{\hbar^{2}}{4m}AA^{\prime\prime}\Psi^{*}\Psi-\frac{\hbar^{2}}{8m}A^{\prime2}\Psi^{*}\Psi+V\left(x\right)\Psi^{*}\Psi.\label{Hamiltonian}
\end{equation}
Having Hamiltonian density, we apply $E=\int\mathcal{H}dx$ to find
the energy of the system. Explicitly, one finds 
\begin{equation}
E=\int\left[\frac{-\hbar^{2}}{2m}\Psi^{*}\left(A^{2}\partial_{x}^{2}+2A^{\prime}A\partial_{x}+\frac{AA^{\prime\prime}}{2}+\frac{A^{\prime2}}{4}\right)\Psi+\Psi^{*}V\left(x\right)\Psi\right]dx\label{Hamiltonian1}
\end{equation}
According to Eq. \eqref{Original-Ham} and Eq. \eqref{Hamiltonian1}
provided that $\Psi\left(x,t\right)=\Psi^{*}\left(x,t\right)$, considering
that $\hat{H}$ is Hermitian the two terms become identical and the
energy reduces to
\begin{equation}
E=\int\Psi\hat{H}\Psi dx=\int\Psi^{\mathcal{*}}\hat{H}\Psi dx=\left\langle \hat{H}\right\rangle ,\label{exp-value}
\end{equation}
in which $\left\langle \hat{H}\right\rangle $ is the expectation
value of $\hat{H}$. Next, we utilize \eqref{Schrodinger-psi} and
\eqref{Schrodinger-phi} to find the continuity equation. Let's multiply
by $\Psi$ and $\Psi^{*}$ from the left the equations \eqref{Schrodinger-phi}
and \eqref{Schrodinger-psi}, respectively. Then by subtraction of
the two equations, we obtain 

\begin{equation}
i\hbar\partial_{t}\left(\Psi\Psi^{*}\right)=-\frac{\hbar^{2}}{2m}\partial_{x}\left(A^{2}\left(\Psi\Psi^{*\prime}-\Psi^{*}\Psi^{\prime}\right)\right).\label{Continuity}
\end{equation}
This is the continuity equation provided we define 
\begin{equation}
\rho=\Psi\Psi^{*},\label{probability density}
\end{equation}
to be the probability density and 
\begin{equation}
J_{x}=\frac{\hbar}{2im}\left(A^{2}\left(\Psi^{*}\Psi^{\prime}-\Psi\Psi^{*\prime}\right)\right),\label{current density}
\end{equation}
 to be the particle current density. Hence the continuity equation
\begin{equation}
\partial_{t}\rho\left(x,t\right)+\partial_{x}j\left(x,t\right)=0\label{ContinuityStandard}
\end{equation}
holds. We would like to comment that the present outcomes in \eqref{probability density}
and \eqref{current density} are significant since the conservation
of probability density is confirmed. 

\section{Lagrangian Density in Complex Domaine : Non-hermitian Schrödinger
Equation}

We expand our discussion into the complex domain such that our proposed
momentum and the potential are complex, $V\left(x\right),A\left(x\right)\in\mathbb{C}$,
provided that both functions be $\mathcal{PT}$-symmetric. Thus, the
Lagrangian density is given by, applying \eqref{generalized-momentum},
\begin{equation}
\mathcal{L}=i\hbar\dot{\Psi}\Psi^{\#}-\frac{\hbar^{2}}{2m}A^{2}\Psi^{\#\prime}\Psi^{\prime}-\left[V\left(x\right)-\frac{\hbar^{2}}{4m}AA^{\prime\prime}-\frac{\hbar^{2}}{8m}A^{\prime2}\right]\Psi\Psi^{\#}\label{Lagrangian-complexPT-1}
\end{equation}
where $\Psi^{\#}=\mathcal{PT}\Psi$. Accordingly, we obtain the Schrödinger
equations using the Euler-Lagrange equations with respect to $\psi^{\#}$
and $\psi$ given by

\begin{equation}
i\hbar\dot{\Psi}=-\frac{\hbar^{2}}{2m}\left(A^{2}\Psi^{\prime}\right)^{\prime}+\left[V\left(x\right)-\frac{\hbar^{2}}{4m}AA^{\prime\prime}-\frac{\hbar^{2}}{8m}A^{\prime2}\right]\Psi\label{Sch-PT-psi-1}
\end{equation}
 and 
\begin{equation}
-i\hbar\dot{\Psi}^{\#}=-\frac{\hbar^{2}}{2m}\left(A^{2}\Psi^{\#\prime}\right)^{\prime}+\left[V\left(x\right)-\frac{\hbar^{2}}{4m}AA^{\prime\prime}-\frac{\hbar^{2}}{8m}A^{\prime2}\right]\Psi^{\#},\label{Sch-PT-psi=000023-1}
\end{equation}
respectively. Moreover, the corresponding Hamiltonian is $\mathcal{PT}$-symmetric
which is admitted in Eq. \eqref{Original-Ham}. Following up the earlier
discussion upon the conservation of probability density, one can approve
the similar method and represent the generalized continuity equation
in
\begin{equation}
\partial_{t}\left(\Psi\Psi^{\#}\right)+\frac{\hbar}{2im}\partial_{x}\left(A^{2}\left(\Psi^{\#}\Psi^{\prime}-\Psi\Psi^{\#\prime}\right)\right)=0.\label{continuity3-1}
\end{equation}
 According to the generalized continuity equation, the probability
density is demonstrated by
\begin{equation}
\rho=\Psi\Psi^{\#}\label{probability-density3-1}
\end{equation}
and the generalized current density admitted in
\begin{equation}
J_{x}=\frac{\hbar}{2im}\left(A^{2}\left(\Psi^{\#}\Psi^{\prime}-\Psi\Psi^{\#\prime}\right)\right).\label{generlized-current-density-1}
\end{equation}
We note that with $\Psi\left(x,t\right)=\Psi^{\#}\left(x,t\right)$,
the particle current density vanishes and consequently, 
\begin{equation}
\frac{d}{dt}\int dx\rho=0,\label{conservation-1}
\end{equation}
which is the conservation of the total probability on $x\in\mathbb{R}$
\cite{Bagchi}. Having been obtained the Hamiltonian density in accordance
with the recent Lagrangian density in 
\begin{equation}
\mathcal{H}=\frac{\hbar^{2}}{2m}A^{2}\Psi^{\#\prime}\Psi^{\prime}-\frac{\hbar^{2}}{4m}AA^{\prime\prime}\Psi^{\#}\Psi-\frac{\hbar^{2}}{8m}A^{\prime2}\psi^{\#}\psi+V\left(x\right)\psi^{\#}\psi,\label{Hamiltonian-den1}
\end{equation}
where the energy is presented by $E=\int\mathcal{H}dx$ related to
the $\mathcal{PT}$-symmetric field theory found to be
\begin{equation}
E=\int\Psi^{\#}\left[\frac{-\hbar^{2}}{2m}\left(A^{2}\partial_{x}^{2}+2A^{\prime}A\partial_{x}+\frac{AA^{\prime\prime}}{2}+\frac{A^{\prime2}}{4}\right)+V\left(x\right)\right]\Psi dx.\label{energy-expectation}
\end{equation}
With the situation that $\Psi\left(x,t\right)=\Psi^{\#}\left(x,t\right)$,
the energy expectation value expressed by
\begin{equation}
E=\int\Psi\hat{H}\Psi dx=\int\Psi^{\#}\hat{H}\Psi dx=\left\langle \hat{H}\right\rangle .\label{energy-expectationPT}
\end{equation}

Now, we suppose that there exsist a second field, according to Ref.
\cite{Mazhari}, defined as $\Psi\left(x,t\right)=\frac{\Phi\left(x,t\right)}{\sqrt{A}}$
and input into Eq. \eqref{Lagrangian-complexPT-1} in which $\Phi\left(x,t\right)$
is a $\mathcal{PT}$-symmetric field, then we find the Lagrangian
density
\begin{equation}
\mathcal{L}=i\hbar\frac{\dot{\Phi}\Phi^{\#}}{A}-\frac{\hbar^{2}}{2m}\left[A\Phi^{\#\prime}\Phi^{\prime}-\frac{1}{2}A^{\prime}\left(\Phi\Phi^{\#}\right)^{\prime}\right]-\left[V\left(x\right)-\frac{\hbar^{2}}{4m}AA^{\prime\prime}\right]\frac{\Phi\Phi^{\#}}{A}.\label{Lagrangian=000023}
\end{equation}
The time-independent Schrödinger equation using the Euler-Lagrange
equations is found 

\begin{equation}
i\hbar\dot{\Phi}=-\frac{\hbar^{2}}{2m}A\left(A\Phi^{\prime}\right)^{\prime}+V\left(x\right)\Phi\label{Sch-PT-psi}
\end{equation}
 and 
\begin{equation}
-i\hbar\dot{\Phi}^{\#}=-\frac{\hbar^{2}}{2m}A\left(A\Psi^{\#\prime}\right)^{\prime}+V\left(x\right)\Phi^{\#},\label{Sch-PT-psi=000023}
\end{equation}
respectively. We multiply Eq. \eqref{Sch-PT-psi} by $\Phi^{\#}$
and Eq. \eqref{Sch-PT-psi=000023} by $\Phi$ then subtract them to
find the generalized continuity equation, and accordingly the probability
density 
\begin{equation}
\rho=\frac{\Phi\Phi^{\#}}{A},\label{probability-density3}
\end{equation}
and the generalized current density 
\begin{equation}
J_{x}=\frac{\hbar A}{2im}\left(\Phi^{\#}\Phi^{\prime}-\Phi\Phi^{\#\prime}\right).\label{generlized-current-density}
\end{equation}
Hence, for the non-Hermitian field which is $\mathcal{PT}$-symmetric
the probability density is conserved. 

\section{Conclusion}

We have introduced a Lagrangian density in field theory for a quantum
particle with a generalized momentum operator \cite{M.H1,M.H2}. By
virtue of the variation of the Lagrangian density, we obtained the
Schrödinger equation which describes the behavior of such quantum
particles in Hermitian and non-Hermitian systems. Furthermore, we
calculated the Hamiltonian density and showed that the energy of the
system is the expectation value of the Hamiltonian operator. Besides,
the obtained outcomes are significant due to the demonstration of
the similar pattern according to the real and complex domains in which
the continuity equation is approved for Hermitian and non-Hermitian
Lagrangian density. Regarding the non-Hermitian case, the continuity
equation is confirmed and shown that it has no contradiction defining
one field according to Ref. \cite{Mazhari} for a complex non-Hermitian
system.

\end{document}